# One pot chemical vapor deposition of high optical quality large area monolayer Janus transition metal dichalcogenides

Ziyang Gan[1]†, Ioannis Paradisanos[2]†, Ana Estrada-Real[2]†, Julian Picker[1], Emad Najafidehaghani[1], Francis Davies[3], Christof Neumann[1], Cedric Robert[2], Peter Wiecha[4], Kenji Watanabe[5], Takashi Taniguchi[6], Xavier Marie[2], Arkady V. Krasheninnikov[3,7], Bernhard Urbaszek[2]*, Antony George[1,8]*, Andrey Turchanin[1,8]*

**Affiliations:**

[1]Institute of Physical Chemistry, Friedrich Schiller University Jena; 07743 Jena, Germany.

[2]Université de Toulouse, INSA-CNRS-UPS, LPCNO; 31077 Toulouse, France.

[3]Institute of Ion Beam Physics and Materials Research, Helmholtz-Centre Dresden-Rossendorf; 01328 Dresden, Germany.

[4]LAAS-CNRS, Université de Toulouse; Toulouse, France.

[5]Research Center for Functional Materials, National Institute for Materials Science; Tsukuba 305-0044, Japan.

[6]International Center for Materials Nanoarchitectonics, National Institute for Materials Science; Tsukuba 305-0044, Japan.

[7]Department of Applied Physics, Aalto University; 00076 Aalto, Finland.

[8]Abbe Center of Photonics; 07745 Jena, Germany.

*Corresponding authors. Email: antony.george@uni-jena.de, urbaszek@insa-toulouse.fr, andrey.turchanin@uni-jena.de.

†These authors contributed equally to this work.

**Abstract:** Scalable growth of engineered two-dimensional (2D) transition metal dichalcogenide (TMD) monolayers provides the opportunity to harness their exotic physical properties for electronic, photonic and quantum applications. Here we report, a simple and reproducible approach which utilizes the catalytic surface characteristics of gold substrate to synthesize 2D monolayers with asymmetric crystal structure, known as Janus TMDs. Gold surface facilitated the diffusion of S atoms underneath chemical vapor deposited $MoSe_2$ monolayers. Then, in an exchange reaction the bottom Se layer is replaced with S atoms at an optimum temperate of 700 °C. The high optical quality of the grown Janus TMD is revealed by optical and magneto-optical spectroscopy performed at cryogenic temperatures.

The physical properties of monolayer (ML) transition metal dichalcogenides (TMD) can be tailored for specific applications by stacking them in heterostructures(*1, 2*), alloying(*3*) or more recently by preparing ML Janus TMDs(*4-6*). Due to the difference in electronegativity of the top and bottom chalcogen layers, Janus TMDs possess an in-built electric dipole that breaks the in-plane mirror symmetry, leading to a manifold of novel physical phenomena such as, *e.g.*, strong Rashba splitting(*7*), piezoelectric(*8*), pyroelectric(*9*), novel-excitonic(*10, 11*) and valleytronic phenomena(*12*). However, the existing synthetic routes for ML Janus TMDs, which depends on the plasma induced replacement of the top chalcogen layer of a parent TMD, are rather complex and it remains challenging to obtain samples with high structural quality over a large surface area (*4, 12-14*). To further realize the potential of Janus TMDs for science and applications, technologically relevant methods for their fabrication still need to be established.

Here we present a simple, scalable and highly reproducible one pot synthesis of ML Janus SeMoS with high optical quality. It is based on a thermodynamic approach resulting in replacement of the bottom Se layer of ML $MoSe_2$ grown on Au foil with S atoms at ~700 °C. The Au substrate plays a crucial role in the formation of Janus SeMoS by acting as a catalytic substrate which promotes adsorption and dissociation of S clusters. The formation of ML Janus SeMoS is demonstrated by angle-resolved X-ray photoelectron and Raman spectroscopy, and the growth mechanism is confirmed by first principles calculations. Encapsulation in hBN reveals spectrally narrow excitonic emission linewidths below 20 meV at cryogenic temperature. As a result of circularly polarized laser excitation, the free exciton peak shows strong circular polarization up to 25% that persists above liquid nitrogen temperatures while temperature dependent linewidth analysis indicates enhanced exciton-phonon coupling. In applied longitudinal magnetic fields, we obtain an exciton g-factor of -3.3. Clear differences in vibrational and optical properties between ML Janus SeMoS as compared to randomized alloys are highlighted.

The synthesis method of ML Janus SeMoS is schematically illustrated in Fig. 1A. Our growth procedure begins with the CVD growth of ML $MoSe_2$ single crystals(*15, 16*). Subsequently, the sample is cooled down to room temperature (RT). The conversion of the $MoSe_2$ into Janus SeMoS takes place upon the temperature increase and annealing at 700 °C in the presence of S vapor. The gold substrate plays an important role in the conversion, as it results in the dissociation of S molecules constituting the vapor phase(*17*) ($S_n$ with n=2, 3, 4, 6, 7 and 8) and chemisorption of the S atoms on the gold substrate(*18*). Upon their diffusion underneath the $MoSe_2$ crystals, they replace the bottom Se layer thereby forming a ML Janus SeMoS. A detailed description of the experimental procedures and a schematic diagram of the CVD process are given in the Supplementary Materials (SM) in fig. S1 and Note 1, respectively. A comparison of our synthetic approach with the previously reported methods is provided in SM Note 2.

The mechanism described above of ML Janus SeMoS synthesis can be understood by analyzing the energetics of the exchange processes in density-functional theory (DFT) calculations(*19*) (see SM Note 3 and fig. S2-S5). We consider the adsorption and migration of S and Se atoms on the Au(111) surface as well as their interaction with the free-standing and supported $MoSe_2$ monolayer. Our calculations show that it is energetically favorable for S molecules to split upon adsorption on Au(111). The formed S adatoms are highly mobile at the temperatures used during the synthesis due to their low migration barriers (< 0.4 eV). Note that it costs more than 5 eV to remove a Se atom from the top layer of the $MoSe_2$ sheet and that splitting of S molecules on this layer is energetically unfavorable. In contrast, the energy penalty for removing a Se atom from the bottom layer of the sheet followed by the immediate adsorption of this atom on the Au surface (outside the ML area) is 1.7 eV only. These calculations clearly support the scenario that during annealing the Se atoms in the bottom layer will predominantly be replaced with S atoms.

In Fig. 1B an optical microscopy image of as-grown ML Janus SeMoS crystals is presented. A high density and homogeneous distribution of the triangular crystals can be clearly recognized. Fig. 1C shows the Janus MLs after their transfer onto a $SiO_2$/Si wafer by electro-chemical bubble transfer technique(*18*). The atomic force microscopy (AFM) topography image of Janus SeMoS on $SiO_2$/Si is provided in Fig. 1D. The AFM height profile (insert to Fig. 1D) and the height distribution histogram (Supporting Information (SI) fig. S6) demonstrate an average thickness of 0.8 ± 0.2 nm as expected for the ML Janus SeMoS(*4*).

To better understand and optimize the conversion process, we performed a series of experiments at different sulfurization temperatures, ranging from 650 °C to 800 °C using the sulfurization time of ~10 min. Afterwards the samples were studied by Raman spectroscopy directly on Au foils, Fig. 1E. The black spectrum shows the Raman signal of as-grown ML $MoSe_2$ with the characteristic $A'_1$ peak at 240 $cm^{-1}$(*20*). When the ML $MoSe_2$ is exposed to S vapour at temperatures ≤ 650 °C (red spectrum), the $A'_1$ peak is barely affected indicating there is negligible or no replacement of the Se. When the sulfurization temperature is optimized to 700 °C (blue spectrum), two characteristic Raman peaks at 290 $cm^{-1}$ ($A^1_1$) and 353 $cm^{-1}$ ($E^2$) are observed confirming the formation of a ML Janus SeMoS(*21*). The narrow full width at half maximum (FWHM) of the $A^1_1$ peak (4.5± 0.2 cm$^{-1}$) indicates a high crystalline quality of the formed ML Janus SeMoS which is consistent with our low temperature Raman data (Fig. 4C below)(*22*). As shown in fig. S7, the sulfurization at 700 °C with time <10 min resulted in the incomplete conversion (see also SM Note 4). With further increase of temperature (green and purple spectra), significant changes in Raman spectra are observed: diminished peak intensity, peak broadening as well as the presence of additional peaks characteristic for MoS$_2$. This behavior corresponds to the formation of a random MoS$_{2(1-x)}$Se$_{2x}$ alloy(*20, 23*), as expected for replacement with S of both bottom and top Se layers; for temperatures ≥800 °C, the Raman spectra are typical of highly defective ML MoS$_2$(*24*). A comparison of the Raman spectra for as-grown Janus SeMoS MLs acquired directly on Au and after the transfer on SiO$_2$/Si wafers reveals identical spectroscopic characteristics showing that the transfer process can be applied in a non-invasive manner, fig. S8A. A Raman map recorded on the transferred Janus ML demonstrates a uniform intensity distribution, hence proving the material homogeneity, fig. S8B-C. Note that besides the individual Janus ML crystals as shown in Fig. 1, also millimeter-scale homogeneous Janus ML films were synthesized by tuning the growth conditions (see fig. S9 and SM Note 5). Our attempt to convert ML MoSe$_2$ into ML Janus SeMoS on commonly used SiO$_2$/Si substrate under the identical experimental conditions as used on Au foils resulted only in the formation of MoS$_{2(1-x)}$Se$_{2x}$ alloys (see fig. S10 and SM Note 6). This observation further shows the crucial role of the Au surface for the successful conversion.

To further analyze the chemical composition and asymmetric structure of synthesized ML Janus SeMoS, we performed angle-resolved X-ray photoelectron spectroscopy (ARXPS)(*25*) of as-grown samples on Au foils, Fig. 2 and fig. S11,12. From the binding energies (BEs) and intensities of the respective XP Mo 3d, S 2p and Se 3d peaks, the formation of ML Janus SeMoS is clearly confirmed (see Fig 2A, SM Note 7, table S1). To demonstrate the asymmetric structure of Janus SeMoS, the XP spectra were measured at various photoelectron emission angles, θ, as illustrated in the inset of Fig. 2B, taking advantage of the enhanced surface sensitivity for larger θ. For each element (Se, Mo, S and Au) the relative intensity (RI) of the XP components were calculated as a function of θ using the equation RI $= \dfrac{I(x,\theta)/I(x,0°)}{I(Au,\theta)/I(Au,0°)}$. From the variation of the RI values for Se, Mo and S, it is seen that the Se layer is on top, followed by Mo and S at the bottom. Thus, ARXPS confirms the exchange of the bottom Se layer in ML MoSe$_2$ with S atoms, in agreement with the energetics analysis from DFT. The atomic ratio of Se:Mo:S is estimated as 0.9±0.2:1.0:1.6±0.2 indicating an excess amount of S atoms chemisorbed on the Au substrate. After the transfer of the ML Janus SeMoS onto a SiO$_2$/Si substrate the obtained ratio is 0.8±0.2:1.0:1.3±0.2, which corresponds to the nearly ideal stoichiometry and suggests a loss of the excess S upon the transfer.

Optical transition linewidths of CVD grown TMD monolayers (typically 50 – 100 meV) can be considerably reduced by removing the samples from the growth substrate and encapsulating them in high quality hBN(*1, 26, 27*). We have applied this approach to our ML Janus SeMoS as sketched in Fig. 3A, where the bottom (top) hBN layer is typically 50 nm (10 nm) thick. We show in Fig. 3B the photoluminescence (PL) from the ML Janus SeMoS in the temperature range from 5 K to RT (see SM Note 1 for experimental details). At T = 5 K the spectrum is dominated by a spectrally narrow peak (FWHM 18 meV) marked as X, visible up to RT. At low energy centered around 1.5 eV we detect a spectrally broad transition marked L, only visible for temperatures up to 150 K.

In Fig. 3C we plot the emission energy of the X peak (exciton) as a function of temperature. Fitting the characteristic redshift of the bandgap (see SM Note 8) allows us to extract an average phonon energy $\langle \hbar \omega \rangle = (38.9 \pm 0.7)$ meV that we use in the fit of the transition linewidth as a function of temperature.(*28*)

$$\gamma = \gamma(0) + a \cdot T + \frac{\beta}{e^{\langle \hbar \omega \rangle / k_B T} - 1},$$

here, $\gamma(0) = (18.5\pm0.1)$ meV is the linewidth of X at T= 0K, $a = (24\pm5)$ μeV/K is the linear broadening due to acoustic phonons and $\beta = (124\pm2)$ meV is the strength of the phonon coupling. The value of $\beta$ found here is considerably larger than the values reported previously for MoSe$_2$ and MoS$_2$ monolayers(*29, 30*). This value could be evidence of stronger exciton-phonon coupling for Janus layers as compared to TMD monolayers due to the intrinsic electric dipole(*31*).

Janus TMD MLs are expected to possess momentum dependent spin splitting and to obey chiral optical selection rules. In Fig. 3B we detect the PL emission both co- and counter- polarized with respect to the excitation laser polarization. We extract the circular polarization degree $P_c = (I_{\sigma+} - I_{\sigma-})/(I_{\sigma+} + I_{\sigma-})$ of the X emission. For an excitation photon energy lying ~180 meV above the X emission energy, we measure a high polarization of $P_c = 25\%$ for temperatures up to 100 K (Fig. 3D) since chiral selection rules hold for the direct transition at the non-equivalent K-points at the Brillouin zone edge(*4, 32*). We find a drastic drop in $P_c$ for temperatures T > 100 K and at 200 K we find $P_c \sim 0$. For comparison, the PL emission from localized states L is unpolarized over the investigated temperature range in Fig. 3B. It is striking that for the X peak the sudden decrease in $P_c$ occurs over the same temperature range that sees drastic changes in the peak position and linewidth plotted in panel 3C, (see SM Note 8 for discussion).

A key feature for ML Janus TMDs is the valley Zeeman splitting of the two exciton states in magnetic fields(6, 33), which was so far not accessible in samples with larger optical transition linewidth. In Fig. 3E we observe a clear energy difference $\Delta Z$ between the $\sigma^+$ and the $\sigma^-$ polarized PL components (see SM Note 8) of ~1.7 meV, at B= 9 T, which corresponds to an exciton Landé g-factor of -3.3. Interestingly, this value lies in between the Landé g-factor of the A-exciton in $MoS_2$, which is reported to be between -2 to -4, depending on background doping(6) and the A-exciton g-factor of $MoSe_2$ of -4(33). In Fig. 3E, the PL intensity is higher from the lower Zeeman branch, which indicates a relaxation towards the energetically lower valley branch during the exciton lifetime.

In Fig. 4A we show power dependent PL spectra using a HeNe laser ($\lambda$=633 nm), and in Fig. 4B we plot the integrated PL intensity $I_{PL}$ as a function of laser power $P$. The X peak intensity increases linearly with power, which is the signature of a free exciton peak, whereas for the L transition the contributing states become saturated(34), see SM Note 8. In addition to our data on the energy shift and broadening with temperature this behavior indicates that the X emission is indeed of excitonic origin. In reflectivity measurements, presented in the same panel, we distinguish the A- and B exciton resonance, marked X and $X_B$, respectively. The difference $\Delta_{X_B-X} \approx 177$ meV, lies between the values reported for $MoS_2$ and $MoSe_2$ monolayers(35).

At the high energy part of the PL spectrum in Fig. 4A close to 1.96 eV, sharp peaks corresponding to Raman scattering signals can be detected. For better visibility and comparison, we plot these spectra in Fig. 4C in units of positive cm$^{-1}$ (Stokes shift). Our data shows the main Raman modes reported for Janus SeMoS(21) in both co- and cross- polarized detection under linear excitation and several additional, yet to be identified features. We note that the Raman peaks at ~155 cm$^{-1}$ and ~175 cm$^{-1}$, linked to the presence of defects(21), are comparatively weak for our samples, which is an indication of high structural quality.

We also plot in Fig. 4C the measurements in a random alloy MoSSe monolayer (hBN encapsulated) measured under the same conditions. The Raman spectra of the random alloy and the ordered Janus layer are clearly different. For the alloy we find the $MoS_2$-like doublet at 400 cm$^{-1}$ reported before in the literature(20, 23). Interestingly, the average phonon energy $\langle \hbar\omega \rangle = (38.9 \pm 0.7)$ meV (corresponds to ~313 cm$^{-1}$) extracted in temperature dependent measurements in Fig. 3C, D falls within the range of the main phonon energies we find in Raman spectroscopy. For further information on the electronic structure, we present in Fig. 4D and 4E PL measurements as a function of excitation laser energy (PLE). The data reveals a strong resonance in absorption at around 1.965 eV. By comparing with reflectivity measurements (Fig. 4A), we ascribe this resonance to the B-exciton of the Janus SeMoS monolayer. Our experiments reveal that the HeNe laser excitation (633 nm; 1.96 eV) is nearly resonant with the B-exciton at T = 4K, which explains the comparatively high PL signal and the very rich structure in the Raman data as compared to 532 nm excitation. We plot for comparison the PLE data for the alloy monolayer, for which we find a resonance at a very similar energy position, but the PLE spectrum at higher energies (>1.97 eV) is very different compared to the Janus monolayer due to the different band structure.

In summary, we developed a simple, reproducible one pot CVD synthesis of large area Janus SeMoS MLs on Au foils. Their high optical quality is confirmed by a spectrally narrow excitonic emission from the Janus SeMoS MLs encapsulated in hBN. The free exciton peak shows strong circular polarization up to 25 % and we extract in applied longitudinal magnetic fields an exciton valley g-factor of -3.3. Temperature dependent linewidth analysis indicates enhanced exciton-phonon coupling.

**Acknowledgments:** The authors thank Stephanie Höppener and Ulrich S. Schubert for enabling the Raman spectroscopy and microscopy studies at the Jena Center for Soft Matter (JCSM).

**Funding:** Jena group received financial support of the Deutsche Forschungsgemeinschaft (DFG) through a research infrastructure grant INST 275/257-1 FUGG, CRC 1375 NOA (Project B2), SPP2244 (Project TU149/13-1) as well as DFG grant TU149/16-1. This project has also received funding from the joint European Union's Horizon 2020 and DFG research and innovation programme FLAG-ERA under grant TU149/9-1. The authors from Toulouse recieved funding from the Institute for Quantum Technologies Occitanie, ANR IXTASE and the Institut Universitaire de France. Growth of hexagonal boron nitride crystals was supported by the Elemental Strategy Initiative conducted by the MEXT, Japan, Grant Number JPMXP0112101001, JSPS KAKENHI Grant Number JP20H00354 and the CREST(JPMJCR15F3), JST. AVK further thanks DFG for the support through Project KR 4866/2-1 and the Collaborative Research Center "Chemistry of Synthetic 2D Materials" SFB-1415-417590517. We also thank the HZDR Computing Center, HLRS, Stuttgart, Germany, and TU Dresden Cluster "Taurus" for generous grants of CPU time.

**Author contributions:** ZG, AG and AT designed the synthesis of Janus TMDs. AT directed the research. ZG synthesized Janus TMDs and performed basic material characterizations. EN and ZG performed the Raman spectroscopy and analysis. JP and CN performed AR-XPS and analysis. FD and AVK performed the DFT calculations and contributed to the interpretation of experimental results. IP and AER performed encapsulation, temperature-dependent optical spectroscopy experiments and spectral analysis. BU, CR, PW and XM designed the low-temperature optical spectroscopy experiments. KW and TT provided the hBN. ZG, AG, I P, BU and AT wrote the manuscript with input from all co-authors.

**Competing interests:** Authors declare that they have no competing interests.

**Data and materials availability:** The data supporting the findings of this study are available from the corresponding authors upon reasonable request.

**Supplementary Materials**

Supplementary Note 1: Methods

Supplementary Note 2: Comparison with previous reports on Janus TMD synthesis

Supplementary Note 3: Detailed explanation of DFT calculations confirming the growth mechanism

Supplementary Note 4: Influence of sulfurization time on the formation of ML Janus SeMoS

Supplementary Note 5: Growth of millimeter scale ML Janus SeMoS

Supplementary Note 6: Growth attempt of ML Janus SeMoS on SiO$_2$/Si

Supplementary Information Note 7: Angle-resolved X-ray photoelectron spectroscopy

Tables S1 to S3

Supplementary information Note 8: Magneto-optical measurements

Figs. S1 to S12

References (*37–56*)

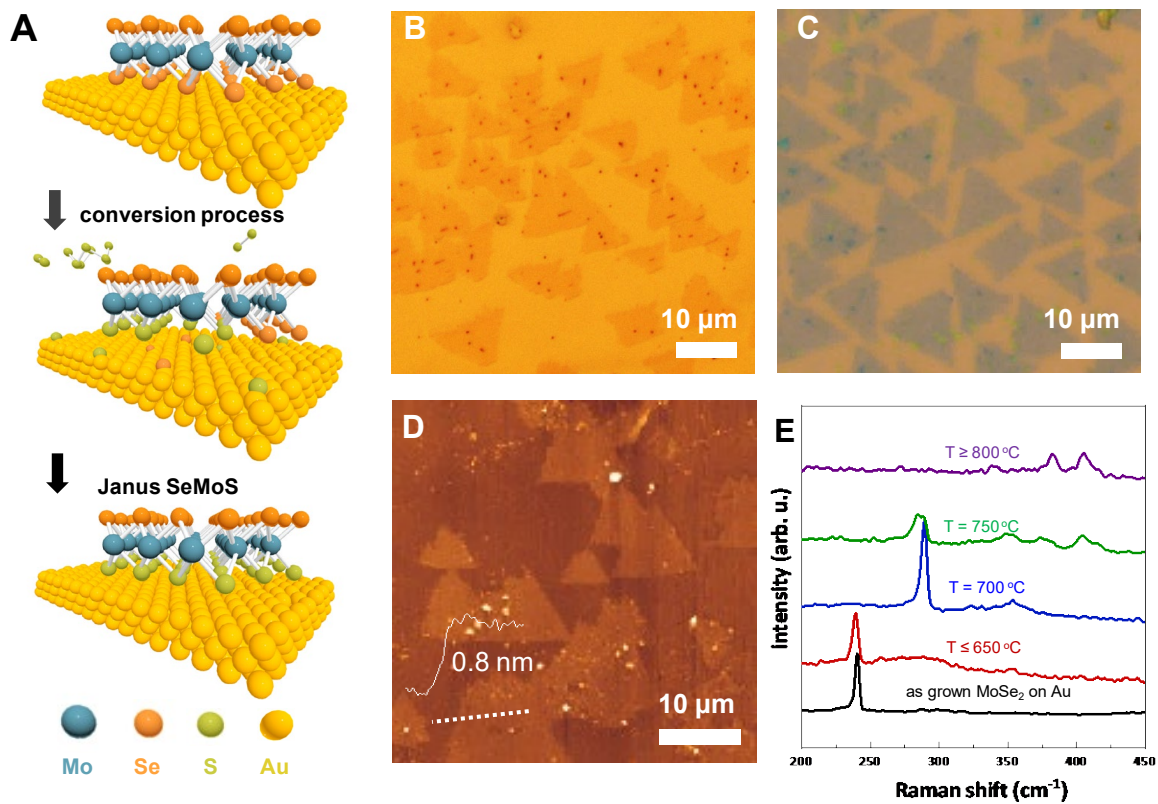

**Fig. 1. Synthesis and microscopic characterization of ML Janus SeMoS. (A)** Schematic illustration of the process for synthesizing ML Janus SeMoS. Initially, ML MoSe$_2$ is synthesized on Au foils by CVD, followed by a conversion process in which the MoSe$_2$ crystals were exposed to S vapour at 700 °C, replacing the bottom-layer Se atoms of MoSe$_2$, which is in contact with the Au surface, by S atoms to form the ML Janus SeMoS. The Au surface absorbs and catalytically dissociate the S clusters. The dissociated S atoms are mobile on the Au surface at high temperature which migrate underneath the ML MoSe$_2$ and exchange with the Se atoms in the bottom layer. On the other hand, replacement of the top Se layer is not energetically favourable, as the S clusters elastically bounced back after colliding with the top Se layer, rather than dissociating and reacting. Optical microscopy images of ML Janus SeMoS, **(B)** as grown on Au foil, and **(C)** transferred on to SiO$_2$/Si substrate. **(D)** AFM topography image of transferred ML Janus SeMoS. The thickness of the ML Janus SeMoS is estimated as 0.8±0.2 nm from the height profile shown in the inset. **(E)** Raman spectra recorded at room temperature using 532nm excitation wavelength on pristine ML MoSe$_2$ and ML MoSe$_2$ exposed to S vapour at different temperatures. The optimum sulfurization temperature for obtaining Janus SeMoS is 700 °C.

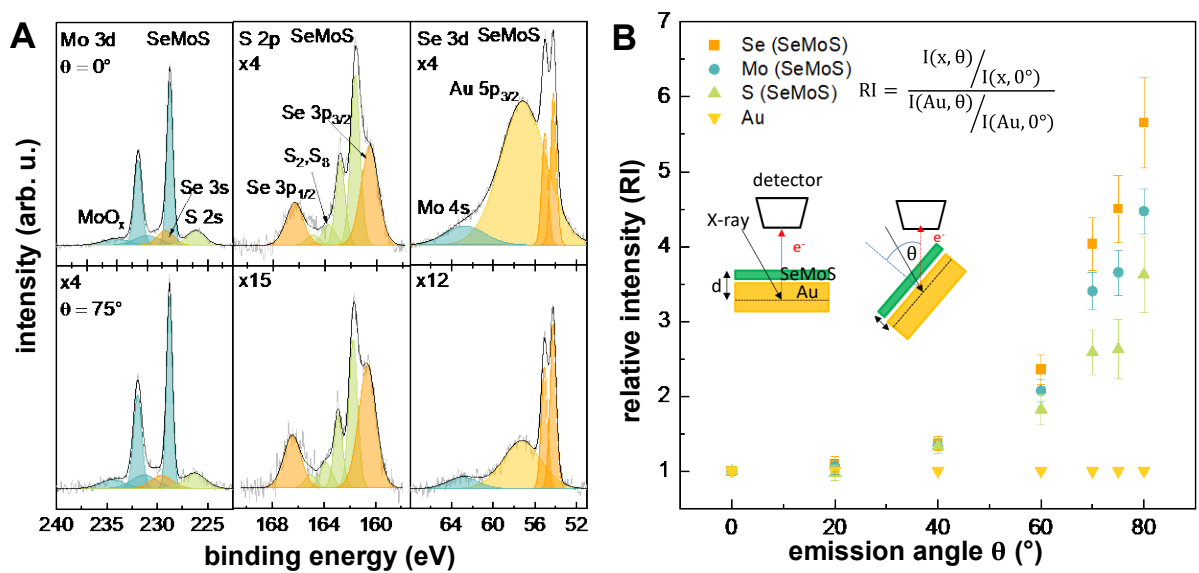

**Fig. 2. Angle resolved X-ray photoelectron spectroscopy of ML Janus SeMoS. (A)** High-resolution Mo 3d, S 2p, and Se 3d XP spectra of as-grown ML Janus SeMoS on Au foil measured at an emission angle (θ) of 0° (normal emission, top) and 75° (bottom), respectively. The fitted components have been named in the figures and discussed in detail in SM Note 7. For better representation, the intensities of the spectra were multiplied by the factors represented in the respective figures. **(B)** Relative intensities (RIs) of the Janus SeMoS components represented by Mo 3d, S 2p and Se 3d peaks as well as the substrate reference Au 4f peak calculated according to the formula written in the inset are plotted for certain emission angles. Thereby, I(x, θ), I(x,0°), I(Au,θ), and I(Au,0°) are the intensities of each element (x = Se, Mo, S, Au) at θ and at normal emission as well as the intensities of Au at θ and at normal emission. Also in the inset, two schemes are shown to illustrate the set-up for ARXPS (θ = 0° and 60°), whereby d is the information depth.

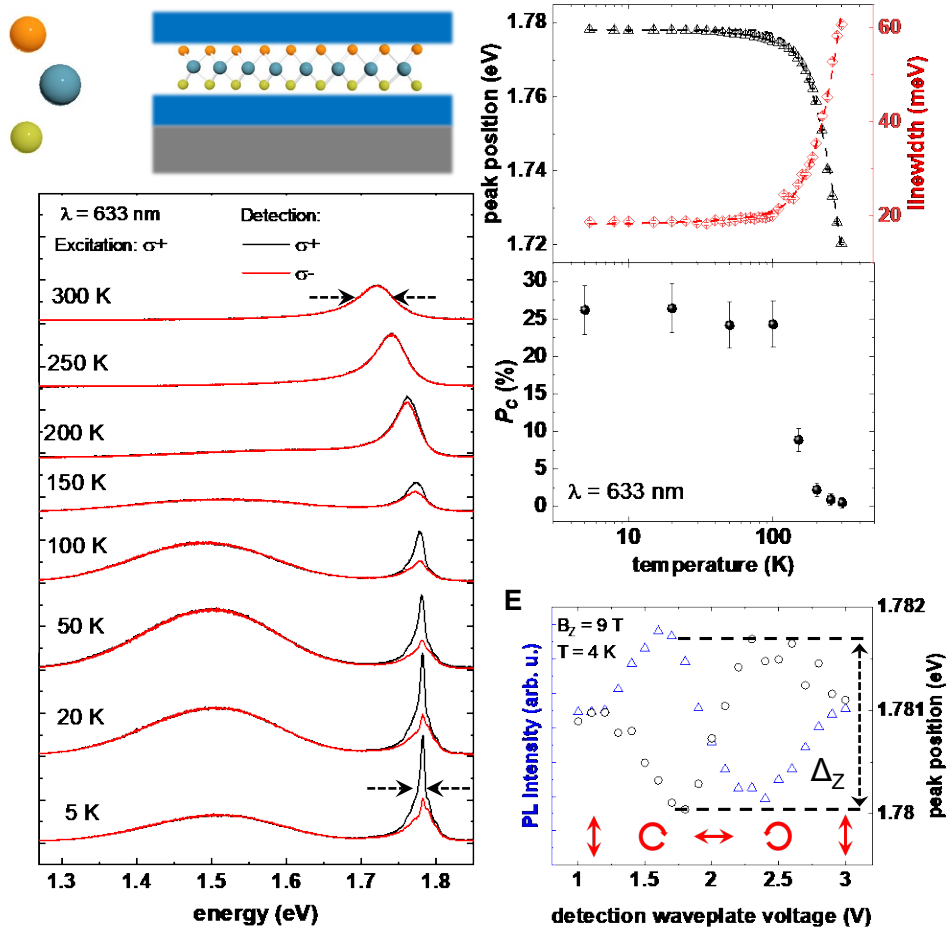

**Fig. 3. Temperature and valley polarization dependent optical properties of hBN encapsulated ML SeMoS.** **(A)** Schematics of the encapsulated Janus SeMoS sample with hBN bottom and top layer. **(B)** Temperature dependent photoluminescence (PL) spectra show the main excitonic emission (X) and up to 150 K also emission from localized states (L). For laser excitation polarization $\sigma^+$ both co-polarized (black) and counter-polarized (red) emission is plotted. **(C)** From measurements as in panel **(B)** both the exciton transition linewidth (right axis, red symbols) and the transition energy (left axis, black symbols) are extracted and fitted, see text. **(D)** The circular polarization degree $P_c = (I_{\sigma+} - I_{\sigma-})/(I_{\sigma+} + I_{\sigma-})$ of the PL from panel **(A)** is plotted as a function of sample temperature. The error bars correspond to the standard deviation extracted over 5 different areas of the sample. **(E)** PL measurements at T=4K in a magnetic field $B_z$= 9T applied perpendicular to the sample surface. The detection polarization is varied by changing the voltage applied to a liquid crystal retarder, which yields the exciton Zeeman splitting $\Delta_Z$. Excitation with linearly polarized light (equal intensity for both circular polarization states). The red arrows indicate the detection polarization state of the liquid crystal retarder for given voltages.

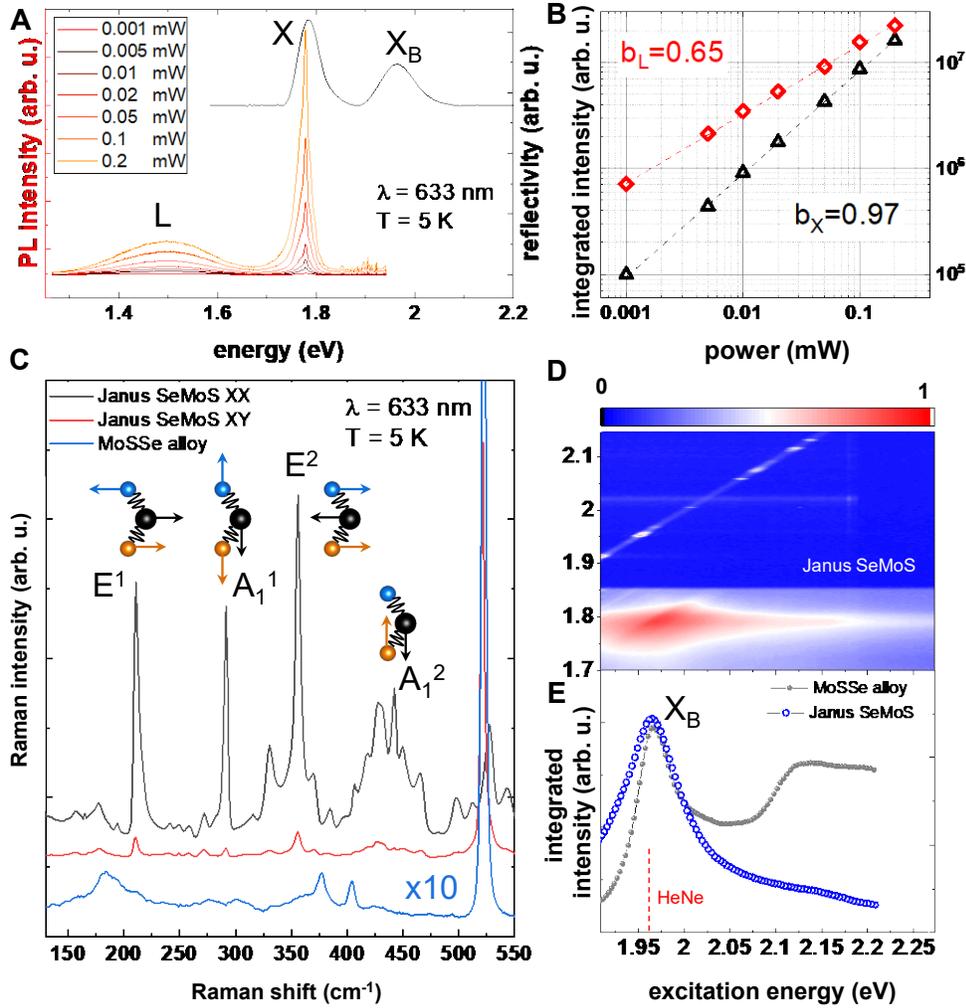

**Fig. 4. Photoluminescence and Raman spectroscopy of hBN encapsulated ML Janus SeMoS.** **(A)** Power dependence of PL emission at T=4K. Superimposed reflectivity shows in addition the B-exciton transition ($X_B$). **(B)** The slopes plotted confirm X as exciton related and L as defect related emission as we fit the data with $I_{PL}(P) = \eta P^b$. **(C)** Comparison of Raman scattering data between SeMoS Janus monolayer (black and red graphs correspond to co- and cross- polarized detection) and MoSSe alloy monolayer (blue graph) which shows $MoS_2$-like doublet around 400 cm$^{-1}$. The main identified modes for the Janus sample are marked. **(D)** In PL excitation spectroscopy the laser excitation energy is varied. The X PL emission is plotted for each value of the excitation energy in a contour plot. The data also shows underlying periodic features that are at a fixed energy with respect to the laser energy, most likely linked to phonon assisted absorption and/or emission(*36*). **(E)** Same experiment as in **(D)** but the integrated X PL emission intensity is plotted for the Janus sample (blue circles) and the alloy monolayer (grey symbols). A strong resonance at the $X_B$ energy is marked.

# Supplementary Materials

# One pot chemical vapor deposition of high optical quality large area monolayer Janus transition metal dichalcogenides

Ziyang Gan[1]†, Ioannis Paradisanos[2]†, Ana Estrada-Real[2]†, Julian Picker[1], Emad Najafidehaghani[1], Francis Davies[3], Christof Neumann[1], Cedric Robert[2], Peter Wiecha[4], Kenji Watanabe[5], Takashi Taniguchi[6], Xavier Marie[2], Arkady V. Krasheninnikov[3,7], Bernhard Urbaszek[2]*, Antony George[1,8]*, Andrey Turchanin[1,8]*

**Affiliations:**

[1]Institute of Physical Chemistry, Friedrich Schiller University Jena; 07743 Jena, Germany.

[2]Université de Toulouse, INSA-CNRS-UPS, LPCNO; 31077 Toulouse, France.

[3]Institute of Ion Beam Physics and Materials Research, Helmholtz-Centre Dresden-Rossendorf; 01328 Dresden, Germany.

[4]LAAS-CNRS, Université de Toulouse; Toulouse, France.

[5]Research Center for Functional Materials, National Institute for Materials Science; Tsukuba 305-0044, Japan.

[6]International Center for Materials Nanoarchitectonics, National Institute for Materials Science; Tsukuba 305-0044, Japan.

[7]Department of Applied Physics, Aalto University; 00076 Aalto, Finland.

[8]Abbe Center of Photonics; 07745 Jena, Germany.

*Corresponding authors. Email: antony.george@uni-jena.de, urbaszek@insa-toulouse.fr, andrey.turchanin@uni-jena.de.

†These authors contributed equally to this work.

**Supplementary Note 1: Methods**

**Synthesis of monolayer (ML) Janus SeMoS on Au foils**

Polycrystalline Au foils (99.95 %, 250 µm thick, 1 × 1 cm$^2$ size, Alfa Aesar) were used as growth substrates. They were initially ultrasonically cleaned in HCl solution (20 wt. %) for 15 min, followed by an acetone solution for 15 min. Then the cleaned Au foils were loaded into the CVD chamber and annealed at 1040 °C for over 10 h under the protection of the mixture gas of Ar/H$_2$ (200/10 cm³/min) at ambient pressure. The annealing treatment reduces the surface roughness of Au foils and promotes their recrystallization toward (111) surface orientation (*37*), which is an epitaxial surface for TMD growth (*37, 38*).

The growth of ML Janus SeMoS was carried out in a three-zone split tube furnace (Carbolite Gero) with a quartz reaction tube of 60 mm diameter. An inner tube with a diameter of 15 mm was used to place the precursors and substrates (Fig. S1). The three-zone configuration allowed to heat the Au foils together with the transition metal oxide precursor and chalcogen precursors (S and Se) individually. Knudsen-type effusion cells were used to deliver the chalcogen precursors (*15,16*). As shown in Fig. S1, the Knudsen cell loaded with S powder (99.98%, Sigma Aldrich) was placed on the first zone of the furnace (Zone 1), another Knudsen cell loaded with Se pellets (99.98%, Sigma Aldrich) was placed on the second zone of the furnace (Zone 2). At the third zone of the furnace (Zone 3), approximately 1-5 µg of MoO$_3$ were placed on a SiO$_2$/Si wafer and the Au foil was placed at the downstream side, next to the wafer. In every growth experiment, together with the Au foils, we have placed a piece of Si wafer with 300 nm thermal oxide layer (Siegert Wafer) to investigate the substrate influence. Next, the quartz tube was evacuated to 1 × 10$^{-2}$ mbar pressure and refilled with argon (5.0, Linde). The whole growth process was carried out under an argon flow of 100 cm³/min and ambient pressure. The argon gas flow was used to carry the chalcogen atoms to the high temperature reaction area where the growth substrates (Au foils and SiO$_2$/Si) were located. A two-step heating protocol was used for subsequent growth of MoSe$_2$, and conversion of Janus structure as explained step by step below.

*Step 1: CVD synthesis of ML MoSe$_2$ on Au Foils:* Initially, the third zone (Zone 3) containing the metal oxides precursor (MoO$_3$) and the Au foils is heated to the growth temperature of 720 °C at a rate of 40 °C/min and held at that temperature for 10 to 30 minutes. The temperature of the second zone (Zone 2) with selenium (Se) was adjusted to reach ~400 °C when the temperature of the growth zone (Zone 3) reaches 700 °C. At the same time, the first zone (Zone 1) with sulfur (S) precursor remains at room temperature. When the temperature of the third zone reaches 720 °C, we introduced hydrogen (5.0, Linde) at a flow rate of 5 cm³/min. The hydrogen flow was turned off after the growth, and the furnace was cooled down to 500 °C under an argon flow of 100 cm³/min. Then the body of the split furnace was opened to rapidly cool down the sample to room temperature. This procedure results in the growth of ML MoSe$_2$ on Au surface. Note that the Ar flow was maintained constant during all the steps from the very beginning of the experiment.

*Step 2: Conversion of MoSe$_2$ to Janus SeMoSe:* After the furnace cooled down to room temperature, the first zone with S precursor and the third zone with Au foils with grown MoSe$_2$ crystals were reheated to 200 °C and 700 °C, respectively. The samples were exposed to the S vapor for 10 min at this temperature. The middle zone (Zone 2) was maintained at a temperature 100 °C to prevent condensation of S during transport to Zone 3. After the conversion process, the split tube furnace was opened to rapidly cool the sample to room temperature. Note that the Ar flow of 100 cm$^3$/min is maintained constant during all the steps from the very beginning of the experiment until opening the chamber to prevent any possible degradation when in contact with air/or atmospheric moisture. This procedure results in the successful conversion of ML MoSe$_2$ to ML Janus SeMoS on the Au surface.

**Electrochemical bubbling transfer of Janus SeMoS films from Au foils**

After growth, the ML Janus SeMoS were detached from the Au foils and transferred onto SiO$_2$/Si wafers using an electrochemical bubble (*18, 39*) transfer method. A PMMA layer of 200 nm (950 kDa, Allresist GmbH, AR-P 679.04) was spin coated onto the Au foil with ML Janus SeMoS. Then, applying a constant current of 20 mA, the PMMA/ML Janus SeMoS was gradually separated from the Au foil by H$_2$ microbubbles produced by water electrolysis in a 0.2 M NaOH solution. After that, the PMMA/ML Janus SeMoS was collected by the target SiO$_2$/Si wafers, followed by a typical acetone and isopropyl alcohol bath process to remove the PMMA support. Using the same electrochemical bubbling transfer process, we additionally transferred Janus MLs on SiO$_2$/Si substrates where high-quality hBN flakes (*40*) were exfoliated prior to transfer. A last step included the transfer of a thin, top hBN to encapsulate the Janus MLs for the low-temperature optical spectroscopy experiments.

**Basic characterization of the ML Janus SeMoS**

**Optical microscopy**

The optical microscopy images were taken with a Zeiss Axio Imager Z1.m microscope equipped with a thermoelectrically cooled 3-megapixel CCD camera (Axiocam 503 color) in bright field operation.

**Atomic force microscopy**

The AFM measurements were performed with a Ntegra (NT-MDT) system in tapping mode at ambient conditions using n-doped silicon cantilevers (NSG01, NT-MDT) with resonant frequencies of 87 – 230 kHz and a typical tip radius of < 6 nm.

**Raman spectroscopy**

The Raman spectra and mapping were acquired using a Bruker Senterra spectrometer operated in backscattering mode. Measurements at 532 nm were obtained with a frequency-doubled Nd:YAG Laser, a 50x objective and a thermoelectrically cooled CCD detector. For all spectra the Si peak at 520.7 cm$^{-1}$ was used for peak shift calibration of the instrument. The Raman spectroscopy maps were obtained using a motorized XY stage and analyzed by fitting a Lorentzian function to the data using a LabView script.

**Angle-resolved X-ray photoelectron spectroscopy (ARXPS)**

XPS was performed in an ultra-high vacuum (base pressure 2×10$^{-10}$ mbar) Multiprobe system (Scienta Omicron) using a monochromatized X-ray source (Al K$_\alpha$) and an electron analyzer (Argus CU) with a spectral energy resolution of 0.6 eV. During the ARXPS measurement, the detected emission angle (ϴ) was varied from 0° to 80 whereas the source analyzer angle (54.7°) was fixed. The spectra were calibrated using the Au 4f$_{7/2}$ peak (84.0 eV) or Si 2p peak (SiO$_2$, 103.5 eV), respectively and fitted using Voigt functions (30:70) [for Au 4f: (20:80)] after a linear (S 2p, Se 3d) or Shirley-background (Mo 3d, Au 4f) subtraction. To perform the quantification, the relative sensitivity factors (RSF, Tab. S3) provided by CasaXPS were used.

**DFT Calculations**

The energetics of the atomic configurations relevant to the exchange mechanism was studied using spin-polarized density functional theory (DFT) as implemented in the VASP code (*41,42*). All the calculations were carried out using the PBE exchange-correlation functional (*43*) and the PAW method (*44*). The van der Waals (vdW) interaction was accounted for using the Grimme scheme (*45*). The structures were fully optimized with an energy cut-off of 400 eV and a force tolerance of 0.01 eVA$^{-1}$. The MoSe$_2$/Au (111) system was modelled as a periodic 8×8 MoSe$_2$ by 9×9 Au supercell composed of 324 Au atoms (four closely packed layers) and 64/128 Mo/Se atoms. The supercell size was chosen to match the theoretical values of the unit cell parameters of the materials, giving rise to a strain of less than 1%. A vacuum space of 15 Å in the perpendicular direction was used. The Brillouin zone of the systems was sampled using 2×2×1 Monkhorst-Pack grid. The test calculations with a 4×4×1 grid gave essentially the same results. The nudged elastic band (NEB) method (*46*) was employed to access migration barriers. These calculations were performed using the same 9x9 Au supercell and employed a 2×2×1 Monkhorst-Pack grid with an energy cut-off of 400 eV. A set of 7 images were used to evaluate the migration barrier, however, due to the symmetry of the diffusion pathway this increases to an effective 13 images.

**Low temperature optical spectroscopy**

Two different optical setups were used for the optical spectroscopy experiments.

For the low-temperature PL, Raman, PLE and reflectivity experiments we use a home-built microscope in a liquid helium (LHe) closed-cycle cryostat system (Attocube, Atto700). A HeNe 633 nm laser is used for the PL and Raman experiments, while for the PLE we utilize a supercontinuum white light laser (NKT Photonics) with a tunable high contrast filter that allows selectivity of the excitation wavelengths. A combination of linear polarizers, halfwave and quarter waveplates allowed the control of excitation and detection polarization for the polarization-resolved Raman and circular polarization PL measurements. The light is focused onto the sample using a microscope objective while the temperature of the sample is kept at T = 5 K. Cryogenic nanopositioners (nm steps, mm range) are used to control the position of the sample with respect to the laser beam. The back-reflected light from the sample is dispersed in a 500mm monochromator (ACTON SpectraPro 2500i, Princeton Instruments) with a 150 g/mm grating (600g/mm in the Raman experiments). The spectra are recorded by a liquid-nitrogen cooled charged coupled device (CCD) array. Low temperature reflectivity experiments are performed using a tungsten-halogen lamp as a white-light source with a stabilized power supply, focused initially on a pinhole that is imaged on the sample. The excitation/detection spot diameter is ≈1µm, i.e., smaller than the typical diameter of the sample.

Magneto-PL experiments have been carried out in an ultra-stable confocal microscope (Attocube, Atto1000) with cryogenic nanopositioners at T = 4 K and in magnetic fields up to 9 T in a Faraday geometry. The detection spot diameter is about 700 nm. The sample is excited by a He-Ne laser (1.96 eV) with linear polarization and both circular $\sigma^+$ and $\sigma^-$ polarized PL signals are detected using a liquid crystal retarder and an analyzer in the detection path. The PL emission is dispersed in a monochromator and detected with a Si-CCD camera.

**Supplementary Note 2: Comparison with previous reports on Janus TMD synthesis**

In general, the previous reports on the preparation of Janus TMDs rely on the replacement of uppermost chalcogens induced by plasma-assisted treatment (*4, 13, 14, 22*) (hydrogen or Se plasma). The general idea is to strip away the top atomic layer by plasma and replace with another chalcogen atom. However, this technique needs the use of plasma treatment, which increases the process complexity. Furthermore, the exposure of the starting transition metal dichalcogenide (TMD) crystal to plasma can create additional unwanted defects (*47*) in the crystal structure. Although this challenge may be partly mitigated through careful synthetic design to enable the successful growth of Janus TMDs, it highly relies on the precise control of many growth parameters (sample positioning, plasma power, high-temperature duration time, and many other parameters related to the kinetics of the reaction) (*13, 14, 22*). In addition, an optimized distance must be maintained between the plasma tail and the growth substrate, leading to short working windows (*13, 22*), thus the controllable growth of large-area Janus TMDs using such a strategy is very challenging. On the other hand, our one-pot CVD approach of growing Janus TMDs on the Au surface is rather straight-forward and easy to reproduce with simple CVD equipment.

Another approach reported for the synthesis of Janus TMD is the high temperature (~800 °C) sulfurization of CVD ML $MoSe_2$ grown on $SiO_2$/Si substrate (*5*). However, such high temperature treatment performed on the CVD grown ML $MoSe_2$ crystals result in the material degradation in our experience, especially when using TMD crystals grown in a separate CVD process. We provide our experimental results on our attempt on $SiO_2$/Si in which we used the similar one-pot approach used for the growth of ML Janus SeMoS in Supplementary Note 5.

**Supplementary Note 3: Detailed explanation of DFT calculations confirming the growth mechanism**

To get a microscopic insight into the mechanism of Janus structure formation, we carried out density-functional theory (DFT) calculations aimed at assessing the energetics of various atomic configurations relevant to the synthesis process. Technical details of calculations are given in method section. The adsorption and migration of S and Se atoms on Au (111) surface were considered, as well as their interaction with the free-standing and supported $MoSe_2$ monolayer, as schematically illustrated in Fig. S5. Our calculations indicate that it is energetically favorable to substitute a Se atom with S atom, but the energy difference is small (about 0.3 eV), consistent with previous theoretical and experimental reports on MoSSe alloys (*19*). Then we studied the atom exchange in the combined $MoSe_2$/Au system represented by a 8×8 $MoSe_2$/ 9×9(111)Au supercell, Fig. S2. Various positions of the S atom in the moiré pattern were considered. These calculations gave roughly the same results as for the free-standing material (the energy difference was less than 0.1 eV) for all the positions, even when placed in the upper and bottom layer of the structure. The reason for that is a smaller atomic radius of S atom as compared to that of Se (105 vs 120 pm, see also Fig. S5B), so that the S atom is 'hidden' and does not interact with the Au substrate, giving rise to the same replacement energy for the upper and lower layers of $MoSe_2$. Thus, the imbalance between the concentration of S atoms in the upper and bottom layers should be governed not by the energetics, but kinetics of the process. To address this conjecture, we considered the adsorption and migration of S and Se atoms on Au surface. The calculations indicated that it is energetically favorable for $S_2$/$S_8$ molecules to split upon adsorption (contrary to the adsorption on $MoSe_2$), Fig. S3. Formation of interstitial atoms in the bulk of Au slab was found to be energetically unfavorable, so that the (111) Au surface should be covered with S adatoms. These are highly mobile at the elevated temperatures used in the experiment with low migration barriers (< 0.4 eV) for adatoms (Fig. S4) found by NEB calculations. We note here that it costs more than 5 eV to remove a Se atom from the MoSe$_2$ sheet, and that splitting of S molecules on the sheet is energetically unfavorable, Fig. S5. At the same time, the energy penalty for removing a Se atom from the bottom layer of the sheet followed by the immediate adsorption of the atom on Au surface (outside the flake area) is 1.7 eV only. Thus, one can expect that during annealing the Se atoms in the bottom layers will predominantly be replaced with S atoms. The scenario is summarized in Fig. S5.

**Supplementary Note 4: Influence of sulfurization time on the formation of ML Janus SeMoS**

We have performed a series of ML Janus SeMoS growth experiments on Au foils in which we changed the sulfurization time to 3, 5 and 10 minutes at 700 °C. The effect of sulfurization time on the Janus conversion process is probed by Raman spectroscopy (Fig. S7). When the sulfurization time is 3 mins, both of MoSe$_2$ peak (A$'_1$, 240 cm$^{-1}$) and Janus SeMoS peak (A$^1_1$, 290 cm$^{-1}$) can be observed in the Raman spectrum, indicated only the bottom Se layer of the ML MoSe$_2$ is partially replaced by S atoms. Furthermore, the intensity of the Janus SeMoS peak (A$^1_1$, 290 cm$^{-1}$) increases with longer conversion time; meanwhile, the intensity of the MoSe$_2$ peak (A$'_1$, 240 cm$^{-1}$) decreases and vanishes after 5 mins. When the sulfurization time is prolonged to 10 mins, the Janus SeMoS signature A$^1_1$ Raman peak at 290 cm$^{-1}$ exhibits a maximized intensity and minimized FWHM, indicating the optimized sulfurization time.

**Supplementary Note 5: Growth of millimeter scale ML Janus SeMoS**

The Au surface possesses the chemical inertness and negligible solubility of metal and chalcogen atoms,[5] allowing the growth of large-area Janus monolayer films. Prolonging the initial MoSe$_2$ growth step to 30 mins resulted in the growth millimeter scale area growth monolayer films on the Au foil (such film areas formed by merging many crystalline grains of MoSe$_2$). Then it is worth directly extending the conversion process to obtain continuous millimeter scale Janus SeMoS films. The Raman spectroscopy analysis performed on the Janus SeMoS films (Fig. S9) shows identical Raman spectra at many different spots with narrow FWHM = 4.6 ± 0.2 cm$^{-1}$ indicating uniformity and high crystalline quality of the Janus SeMoS films (*4, 5, 22*).

**Supplementary Note 6: Growth attempt of ML Janus SeMoS on SiO$_2$/Si**

We performed all the growth and sulfurization processes on SiO$_2$/Si substrate together with the Au foils for direct comparison between the growth on Au foils and SiO$_2$/Si substrates at different temperatures. The Raman spectroscopy data performed on the grown material on SiO$_2$/Si substrate is provided in Fig. S10. As evident from the Raman spectra we could only observe the formation of MoS$_{2(1-x)}$Se$_{2x}$ alloy(*20*) and Janus SeMoS did not occur. The samples sulfurized at 850 °C show characteristics peaks of MoS$_2$ with broader full width at half maximum (FWHM) indicating the conversion to highly defective MoS$_2$ (*48, 49*). In addition, when the sulfurization temperature is below 700 °C, only the MoSe$_2$ signature can be observed, which indicates that S does neither substitute the top and nor bottom Se layer of ML MoSe$_2$ without the catalytic effect of the Au substrate.

**Supplementary Note 7: Angle-resolved X-ray photoelectron spectroscopy**

For the as grown Janus SeMoS sample the main peaks are already discussed in main the paper. From quantification of the SeMoS components in the Mo 3d, S 2p and Se 3d regions we calculate the ratios of Se:Mo:S to be 0.9(2):1.0(2):1.5(2) indicating a S excess. This excess could be explained by S-Au or C-S-Au bounds, which are located at a binding energies (BE) of ≈ 161.3 Ev (*50*) and ≈ 161.9 eV (S $2p_{3/2}$) (*51*), respectively. Thus, it isn't possible with XPS to distinguish S atoms which are bound to the substrate or arranged in the Janus SeMoS structure. For calculation of the relative intensity (RI) according to the equation $RI = \frac{I(x,\theta)/I(x,0°)}{I(Au,\theta)/I(Au,0°)}$ the intensities of the SeMoS peaks in the Mo 3d, S 2p and Se 3d region as well as the Au 4f substrate peaks (Fig. S11) for different emission angles θ are included. The RI values show the θ-dependend intensity change of each element $I(x,\theta)/I(x,0°)$ (x = Se, Mo, S and Au) relative to the change of intensity of the substrate peaks $I(Au,\theta)/I(Au,0°)$. If we take the fact into account that the higher θ the higher the surface sensitivity and the lower the information depth $d = \lambda \cos\theta$ where λ is the attenuation length (*52*), we can directly derive the relative positions of Se-, Mo- and S-atoms in a quantitative manner from Fig. 2B. The highest RI values we obtain for Se followed by Mo and S. Thus, we find from ARXPS measurements that Se is the top layer and S the bottom layer. This result confirms the exchange of the bottom Se layer in $MoSe_2$ with S atoms as described in the main paper. In addition, we could observe further components marked in Fig. 2A. In the Mo 3d region we see a doublet at ≈ 231.0 eV (Mo $3d_{5/2}$) and ≈ 234.2 (Mo $3d_{3/2}$) originating from the precursor $MoO_x$. Also, the S 2s (BE = 226.2(1) eV) and Se 3s (BE = 229.3(2) eV) orbitals are overlapping with this spectrum. Besides the sharp S 2p doublet already discussed, there is a second S doublet observed at higher binding energies (BE[S $2p_{3/2}$] = 163.6(2) eV). These photoelectrons originating from S which is probably arranged in $S_2$ or $S_8$ molecules on the surface. This second S doublet was not considered for the calculation of RI, as well as the $MoO_x$ in the Mo 3d spectra. In the S 2p region also the Se 3p orbital is overlapping. We can separate these peaks from S 2p peaks due to the larger full width of half maximum (FWHM) and larger spin-orbit splitting (*25*). The intensity ratio between Se 3p and S 2p increases for higher emission angle clearly shown in Fig. 2A. Thereby, we can directly see that Se must be higher-lying than S. In the Se 3d region the Se 3d doublet originating from the Janus SeMoS structure is overlapping with the Au $5p_{3/2}$ substrate peak as well as the Mo 4s orbital. Information about the components discussed are summarized in Tab. S1. Further elements on the samples are O and C (Fig. S11) originating mostly from the ambient conditions (or from $MoO_x$). For the transferred Janus SeMoS to $SiO_2$/Si similar SeMoS peaks for each element could be observed (Fig. S12). Whereby, the intensity ratios of Se:Mo:S are calculated to be 0.8(2):1.0(2):1.3(2) which roughly fits the expected structure of 1:1:1. The measured S excess for samples on gold is reduced here because the S bond to the Au substrate is not transferred and just the S in the Janus SeMoS configuration contribute to the main S peaks. Besides this behaviour, also the S 2p peaks originating from the $S_2$ or $S_8$ molecules disappear. Information about the peaks for the Janus SeMoS on $SiO_2$/Si are given in Tab. S2. Note, the discussed transferred Janus SeMoS was a different sample compared to the discussed Janus SeMoS on Au foil.

**Table S1**: Quantitative analysis of the high-resolution XP spectra shown in Fig. 2A of the main paper including the peak assignment, their binding energies, their full width at half maximum

(FWHM) and areas obtained from the spectra deconvolution. For the p and d photoelectron doublets, the fixed intensity ratios of 2:1 as well as 3:2 originating from spin-orbit coupling are used except for the Se 3d orbital. Here, an intensity ratio of 1.3:1 are employed (*53*). The numbers in the parentheses are the experimental error represented as the uncertainty of the last digit.

| Peak Assignment | Binding energy, eV | FWHM, eV | Area, % |
|---|---|---|---|
| *Janus SeMoS on Au foil* $\theta = 0°$ $\theta = 75°$ | | | |
| Mo $3d_{5/2}$ (SeMoS) | 228.7(1) | 0.7(1) | 48(2) |
|  | 228.8(1) | 0.7(1) | 46(2) |
| Mo $3d_{3/2}$ (SeMoS) | 231.9(1) | 0.9(1) | 32(2) |
|  | 231.9(1) | 1.0(1) | 31(2) |
| Mo $3d_{5/2}$ (MoO$_x$) | 231.0(3) | 3.0(3) | 12(2) |
|  | 231.3(3) | 3.0(3) | 14(2) |
| Mo $3d_{3/2}$ (MoO$_x$) | 234.2(3) | 3.0(3) | 8(2) |
|  | 234.4(3) | 3.0(3) | 9(2) |
| Se 3s | 229.2(3) | 1.9(2) | 100 |
|  | 229.6(3) | 2.6(2) | 100 |
| S 2s | 226.2(2) | 2.1(2) | 100 |
|  | 226.3(2) | 2.6(2) | 100 |
| | | | |
| S $2p_{3/2}$ (SeMoS) | 161.6(1) | 0.9(1) | 57(2) |
|  | 161.7(1) | 0.8(1) | 54(2) |
| S $2p_{1/2}$ (SeMoS) | 162.8(1) | 0.9(1) | 28(2) |
|  | 162.9(1) | 0.8(1) | 27(2) |
| S $2p_{3/2}$ (S$_2$, S$_8$) | 163.7(2) | 1.2(2) | 10(2) |
|  | 163.8(2) | 1.1(2) | 13(2) |
| S $2p_{1/2}$ (S$_2$, S$_8$) | 164.9(2) | 1.2(2) | 5(2) |
|  | 165.0(2) | 1.1(2) | 7(2) |
| Se $3p_{3/2}$ | 160.5(1) | 1.6(1) | 67 |
|  | 160.6(1) | 1.6(1) | 67 |
| Se $3p_{1/2}$ | 166.3(1) | 1.6(1) | 33 |
|  | 166.4(1) | 1.6(1) | 33 |
| | | | |
| Se $3d_{5/2}$ (SeMoS) | 54.2(1) | 0.7(1) | 58 |
|  | 54.2(1) | 0.7(1) | 58 |
| Se $3d_{3/2}$ (SeMoS) | 55.0(1) | 0.7(1) | 42 |
|  | 55.0(1) | 0.7(1) | 42 |
| Au $5p_{3/2}$ | 57.1(2) | 5.1(3) | 100 |
|  | 57.2(2) | 4.3(3) | 100 |
| Mo 4s | 62.7(2) | 4.4(2) | 100 |
|  | 62.9(2) | 3.5(2) | 100 |

**Table S2:** Quantitative analysis of the high-resolution XP spectra including the peak assignment, their binding energies, their FWHM and areas obtained from the spectra deconvolution. For the p and d photoelectron doublets, the fixed intensity ratios of 2:1 as well as 3:2 originating from spin-orbit coupling are used except for the Se 3d orbital. Here, an intensity ratio of 1.3:1 are employed (*53*). The numbers in the parentheses are the experimental error represented as the uncertainty of the last digit.

| Peak Assignment | Binding energy, eV | FWHM, eV | Area, % |
|---|---|---|---|
| *Janus SeMoS on SiO$_2$/Si* $\theta = 0°$ $\theta = 75°$ | | | |
| Mo 3d$_{5/2}$ (SeMoS) | 229.0(1) | 0.9(1) | 53(2) |
|  | 228.9(1) | 1.0(1) | 53(2) |
| Mo 3d$_{3/2}$ (SeMoS) | 232.2(1) | 1.0(1) | 36(2) |
|  | 232.1(1) | 1.2(1) | 36(2) |
| Mo 3d$_{5/2}$ (MoO$_x$) | 232.0(3) | 3.0(3) | 7(2) |
|  | 231.5(3) | 2.0(3) | 7(2) |
| Mo 3d$_{3/2}$ (MoO$_x$) | 235.1(3) | 3.0(3) | 4(2) |
|  | 234.6(3) | 2.0(3) | 4(2) |
| Se 3s | 229.3(3) | 2.7(2) | 100 |
|  | 228.9(3) | 3.0(2) | 100 |
| S 2s | 226.4(2) | 1.9(2) | 100 |
|  | 226.2(2) | 1.6(2) | 100 |
| | | | |
| S 2p$_{3/2}$ (SeMoS) | 161.9(1) | 0.8(1) | 67 |
|  | 161.8(1) | 0.9(1) | 67 |
| S 2p$_{1/2}$ (SeMoS) | 163.1(1) | 0.8(1) | 33 |
|  | 163.1(1) | 0.9(1) | 33 |
| Se 3p$_{3/2}$ | 160.7(1) | 1.6(1) | 67 |
|  | 160.6(1) | 1.8(1) | 67 |
| Se 3p$_{1/2}$ | 166.5(1) | 1.6(1) | 33 |
|  | 166.4(1) | 1.8(1) | 33 |
| | | | |
| Se 3d$_{5/2}$ (SeMoS) | 54.4(1) | 0.8(1) | 58 |
|  | 54.3(1) | 1.0(1) | 58 |
| Se 3d$_{3/2}$ (SeMoS) | 55.2(1) | 0.8(1) | 42 |
|  | 55.1(1) | 1.0(1) | 42 |
| Se loss | 59.2(3) | 3.7(3) | - |
|  | - | - | - |
| Mo 4s | 63.6(2) | 2.7(2) | 100 |
|  | 63.4(2) | 2.0(2) | 100 |

**Table S3:** Quantitative analysis of the SeMoS components of the high-resolution Mo 3d, S 2p and Se 3d XP spectra for different emission angles $\theta$. For quantification the relative sensitivity factors (RSF) from CasaXPS depending on the orbital, the analyser and the set-up geometry are used to be able to compare the peak intensities for different elements. These RSF values for Mo $3d_{5/2}$, S $2p_{3/2}$ and Se $3d_{5/2}$ are 5.62, 1.11 and 1.36, respectively. The numbers in the parentheses are the experimental error represented as the uncertainty of the last digit.

| Emission angle $\theta$ | Mo (SeMoS), at% | S (SeMoS), at% | Se (SeMoS), at% |
|---|---|---|---|
| *Janus SeMoS on Au foil* | | | |
| 0° | 28(2) | 45(2) | 27(2) |
| 20° | 29(2) | 43(2) | 28(2) |
| 40° | 28(2) | 45(2) | 27(2) |
| 60° | 29(2) | 40(2) | 31(2) |
| 70° | 30(2) | 36(2) | 34(2) |
| 75° | 30(2) | 35(2) | 35(2) |
| 80° | 29(2) | 36(2) | 35(2) |
| *Janus SeMoS on SiO$_2$/Si* | | | |
| 0° | 32(2) | 43(2) | 24(2) |
| 75° | 32(2) | 38(2) | 30(2) |

**Supplementary information Note 8: Magneto-optical measurements**

We perform magneto-optics measurements (*26*) in a static magnetic field of B = 9 Tesla applied perpendicular to the monolayer (Faraday geometry) in our hBN encapsulated samples. In this perpendicular magnetic field, the degeneracy of the optical transitions in the K$^+$ and K$^-$ valley is lifted as they are energetically separated by the valley Zeeman splitting $\Delta Z = \mu_B g_X B$. In the detection path we use a liquid crystal retarder, that allows to gradually change optical retardation from zero to multiples of the emission wavelength $(0, \frac{\lambda}{4}, \frac{\lambda}{2}, ...)$. As a function of the applied voltage the detected PL is therefore first linear, then elliptical, then circular and then linear again, as sketched in Fig. 3E. The sample is excited with linearly polarized light.

**Optical spectroscopy analysis**
In Fig. 3C we plot the emission energy of the X peak as a function of temperature. We fit the characteristic redshift of the bandgap with an empirical formula that allows us to extract an average phonon energy $\langle \hbar \omega \rangle$ :

$$E_G^T = E_G(0) - S\langle \hbar \omega \rangle \left\{ \coth\left[\frac{\langle \hbar \omega \rangle}{2k_B T}\right] - 1 \right\}$$

here, $E_G(0)$ is the optical gap at T = 0 K, $S$ is a dimensionless coupling constant, $k_B$ is Boltzmann's constant and $\langle \hbar \omega \rangle$ is the average phonon energy. We obtain values of $E_G(0) = (1.778 \pm 0.001)$ eV, $S = 2.4 \pm 0.1$ eV and $\langle \hbar \omega \rangle = (38.9 \pm 0.7)$ meV, in order of magnitude agreement with reported values on Janus monolayers (*28*). Our CVD grown samples allow access to excitonic properties so far mainly reported for Janus samples from transformed exfoliated flakes (*28, 54*).

In Fig. 4A we show power dependent PL measurements using a HeNe laser (λ=633 nm), we plot in Fig. 4B the integrated PL intensity $I_{PL}$ for the L and the X peak as a function of laser power $P$ that we fit with $I_{PL}(P) = \eta P^b$ where $\eta$ is a proportionality factor that depends on generation, emission and detection efficiency of the experiment (*5*). The L peak shows a slope below one with $b_L = 0.65$. This indicates saturation of $I_{PL}$ with increasing power due to the finite number of available defect/localization sites. The X peak intensity increases linearly with power as $b_X = 0.97$, which is the signature of a free exciton peak (*34*).

The exciton polarization dynamics in TMD monolayers is governed by the long-range Coulomb exchange interaction (*55*). Although it is challenging to analyse the polarization dynamics from time-integrated measurements in detail, one possibility is that the increased exciton scattering rate at higher temperature leads to a shortening of the exciton polarization lifetime (*55*), and at the same time the PL emission time increases. This would lead to a lowering of the measured $P_c$. In addition, in our experiment the laser energy is fixed at 1.96 eV, and the X peak energy strongly decreases with temperature. This change in energy difference for the non-resonant excitation can induce changes in the valley polarization generation and relaxation.

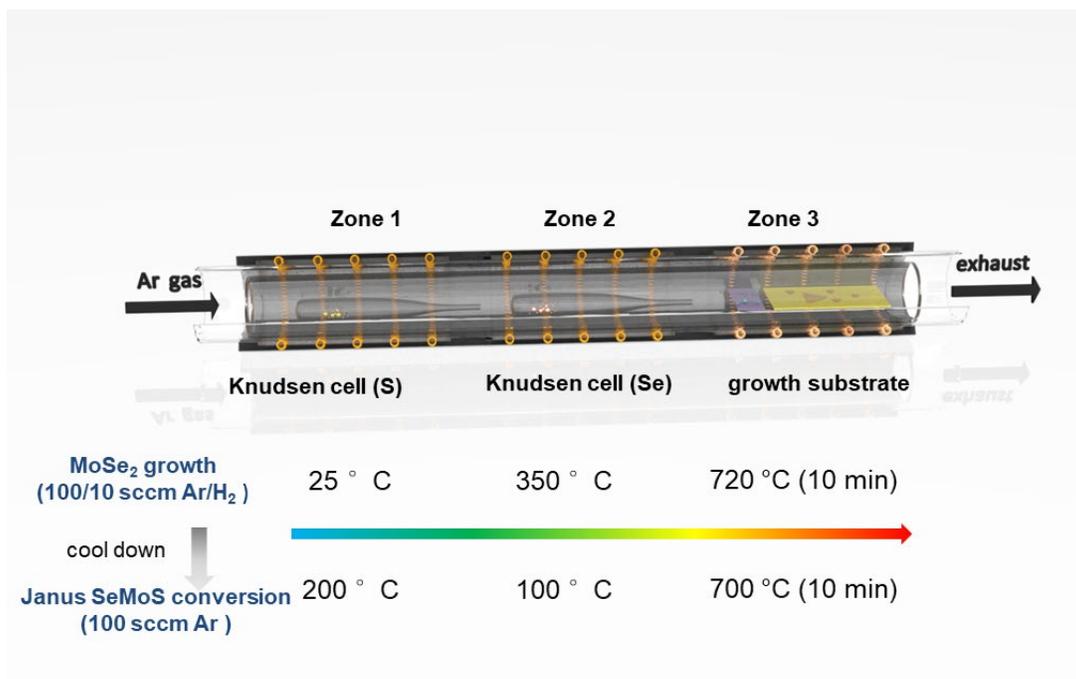

**Fig. S1:** Schematic diagram of the schematic of the one-pot CVD setup with process flow.

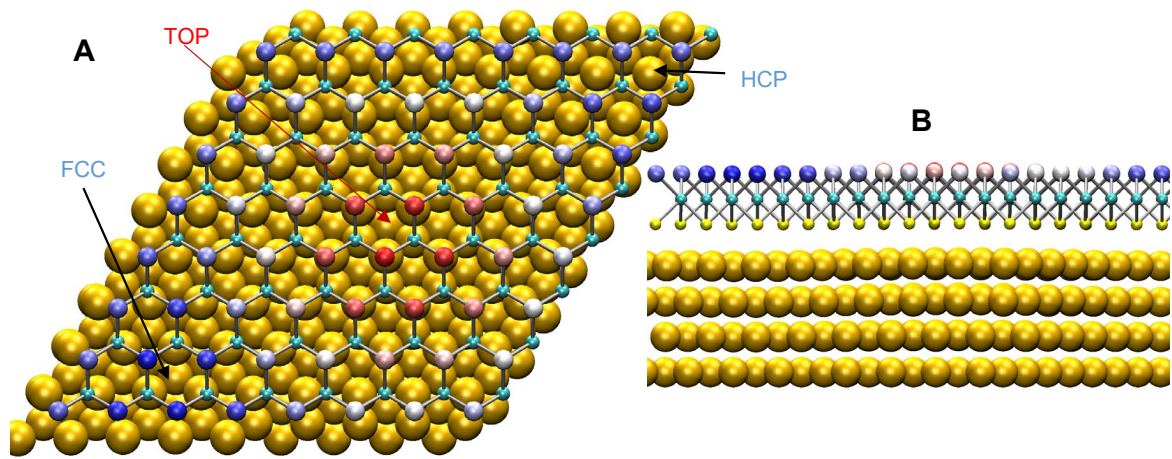

**Fig. S2: (A)** The atomic structure of MoSe$_2$ sheet on Au(111) surface corresponding to a periodic 8×8 MoSe$_2$ by 9×9 Au pattern. The top Se atoms are colored according to their elevation, as also seen from panel **(B)**. It is evident that the bottom Se atoms have different atomic environment, e.g., there is an Au atom exactly under Se atom in the TOP area, while no atom under Se atoms in FCC or HCP regions.

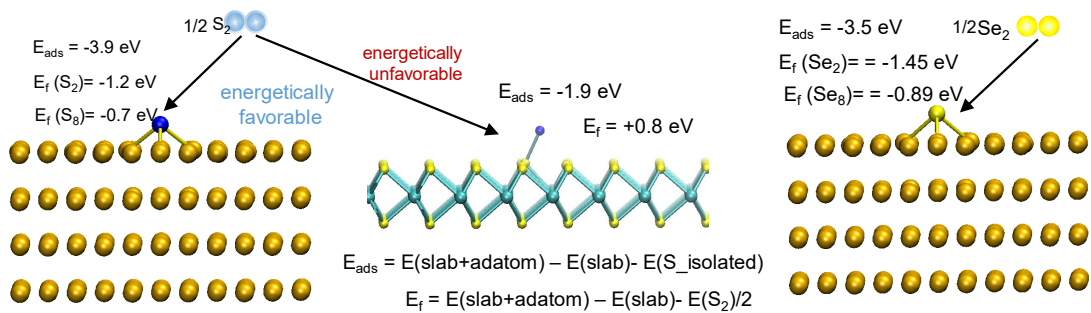

**Fig. S3:** Energetics of S and Se atoms on top of Au (111) and MoSe$_2$. Adsorption energy E$_{ads}$ is the energy gain upon adsorption of isolated atom, and formation energy of adatom $E_f$ is defined as the energy (per atom) required to break up the S/Se molecule and place the atom on the Au surface or on top of a MoSe$_2$ sheet.

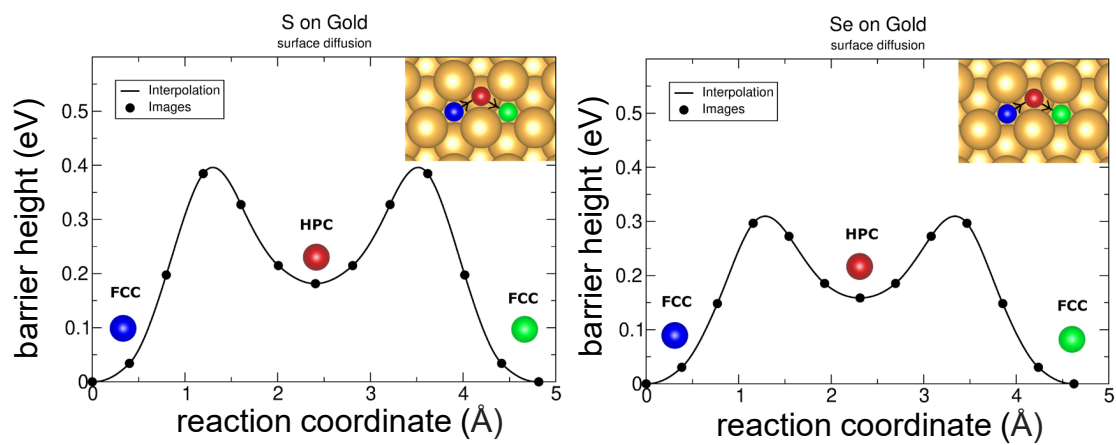

**Fig. S4:** Migration barriers and migration paths of S and Se adatoms on top of Au(111) Au surface as calculated using the NEB method

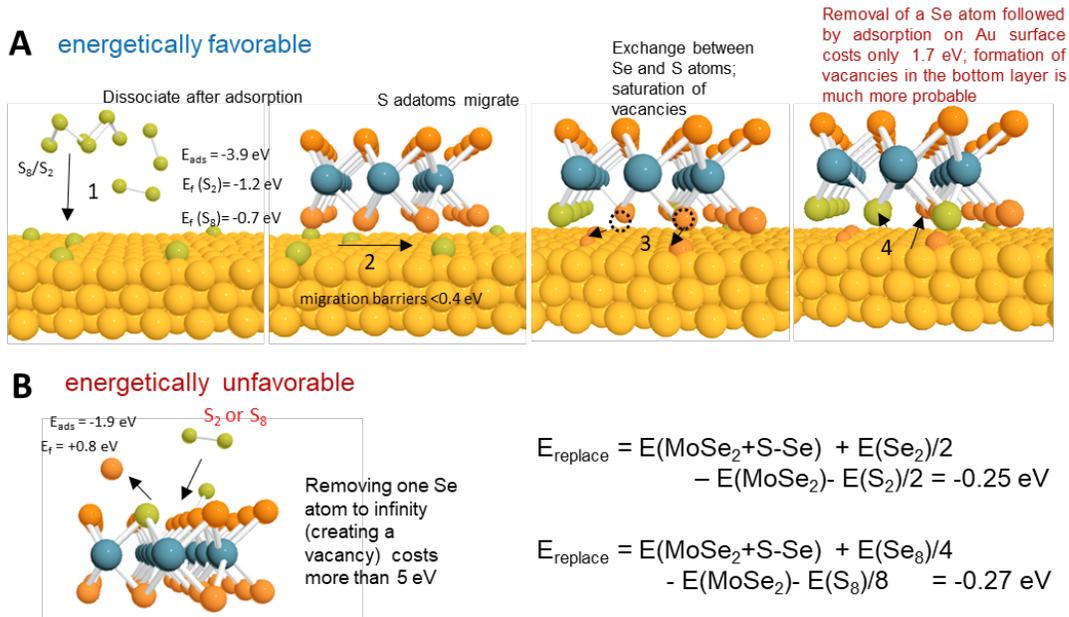

**Fig. S5:** Detailed schematic diagram explaining the conversion mechanism based on replacing bottom Se atoms facing the Au(111) surface with S atoms in S atmosphere as revealed by DFT calculations. **(A)** On Au(111) surface, the $S_2$ dimers can easily dissociate into S adatoms, which can diffuse very easily at the elevated temperatures used in the experiment because of their low migration energy barriers (< 0.4 eV). In addition, removing a Se atom followed by its adsorption on Au surface requires only 1.7 eV, which is much lower than the energy required (> 5 eV) for direct formation of Se vacancy in $MoSe_2$ (taking the energy of an isolated Se atom as a reference). At the same time, dissociation of the $S_2$ dimer on top of $MoSe_2$ monolayer is energetically unfavorable, as illustrated in panel **(B)**. The replacement of Se atom with S is energetically favorable both in the free-standing and supported $MoSe_2$ sheet, but the energy gain is roughly the same for both systems. Moreover, the energy only weakly depends on the side and position of the atom in the Moiré pattern. The reason for that is a smaller atomic radius of S atom as compared to that of Se (105 vs 120 pm), so that the S atom is 'hidden' and does not interact with the Au substrate, giving rise to the same replacement energy for the upper and lower layers of $MoSe_2$. Thus, the imbalance between the concentration of S atoms in the upper and bottom layers should be governed not by the energetics, but kinetics of the process.

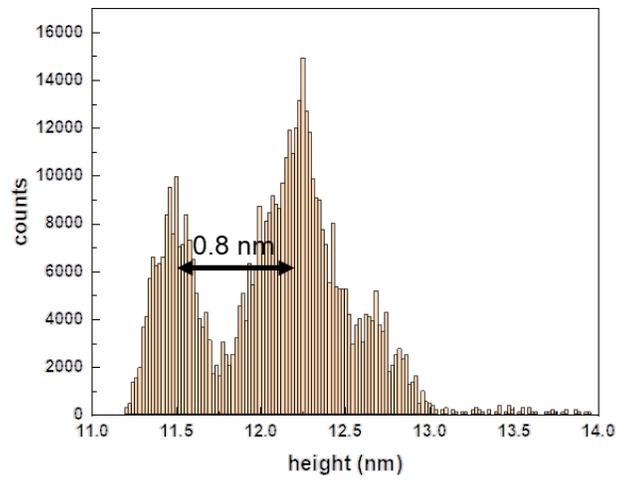

**Fig. S6:** Height distribution histogram between the $SiO_2$ substrate and a monolayer Janus SeMoS crystal (data extracted from Fig. 1D).

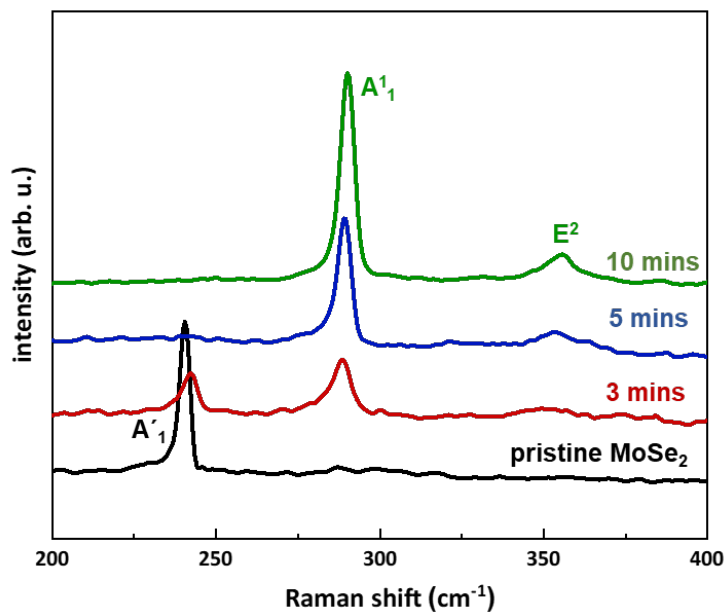

**Fig. S7:** Influence of sulfurization time in the conversion of the monolayer $MoSe_2$ grown on Au foils to Janus SeMoS. In these experiments different set of monolayer $MoSe_2$ crystals grown on Au foils were exposed to S vapor at 700 °C for 3, 5 and 10 minutes. The effect of sulfurization time on the Janus conversion process is then probed by Raman spectroscopy. When the sulfurization time is 3 mins, both of $MoSe_2$ signature ($A'_1$, 240 cm$^{-1}$) and Janus SeMoS signature ($A^1_1$, 290 cm$^{-1}$) can be observed in the Raman spectrum, indicated only partial regions are replaced by S atoms in the bottom Se layer of $MoSe_2$. Furthermore, the intensity of the Janus SeMoS signature ($A^1_1$, 290 cm$^{-1}$) increases with longer conversion time; meanwhile, the intensity of the $MoSe_2$ signature ($A'_1$, 240 cm$^{-1}$) decreases and vanishes in 5 mins. While the sulfurization time prolong to 10 mins, the Janus SeMoS signature $A^1_1$ Raman peak at 290 cm$^{-1}$ exhibits a maximized intensity and minimized FWHM, indicted the optimized sulfurization duration.

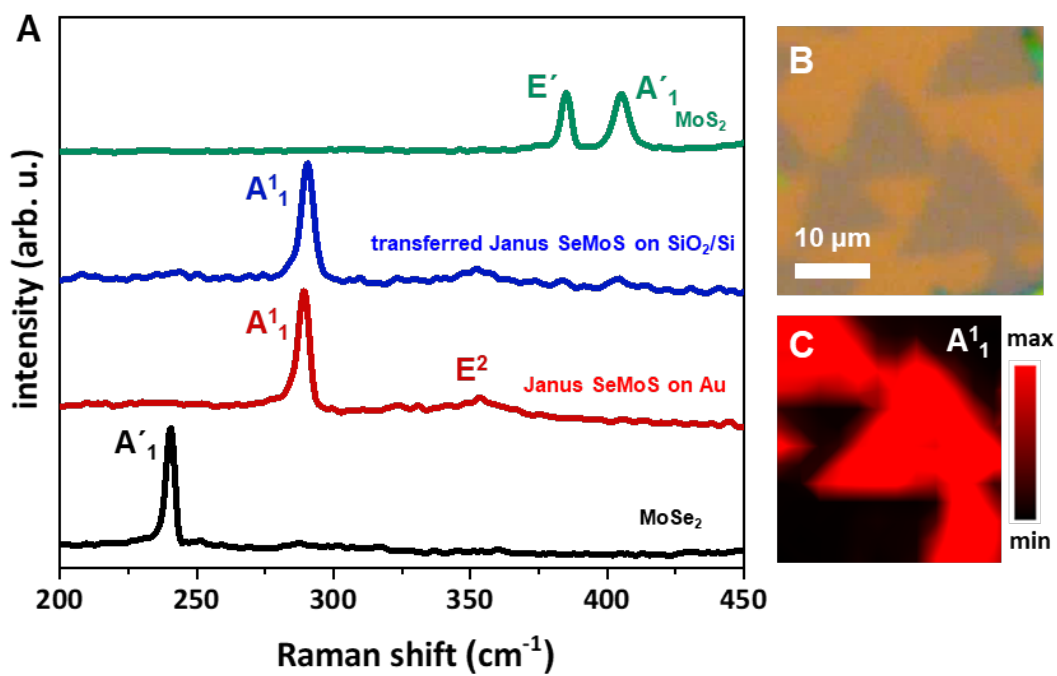

**Fig. S8:** **(A)** Raman spectra of pristine MoSe$_2$ on Au, pristine Janus SeMoS on Au, transferred SeMoS on SiO$_2$/Si on Au and pristine MoS$_2$ on Au. The as grown and transferred Janus SeMoS monolayer crystals show similar Raman spectra resulting from the successful transfer process without changing the crystal quality. **(B)** Optical microscopy image (OM) of transferred isolated Janus SeMoS monolayer crystals on SiO$_2$/Si. The synthesis of isolated Janus SeMoS crystals starts with the growth of MoSe$_2$ on the Au foil at ~720 °C for ~10min. **(C)** Raman intensity mapping performed from **(B)**. The $A^1_1$ mode, originating from the out-of-plane vibration of S-Mo-Se bonds is mapped.

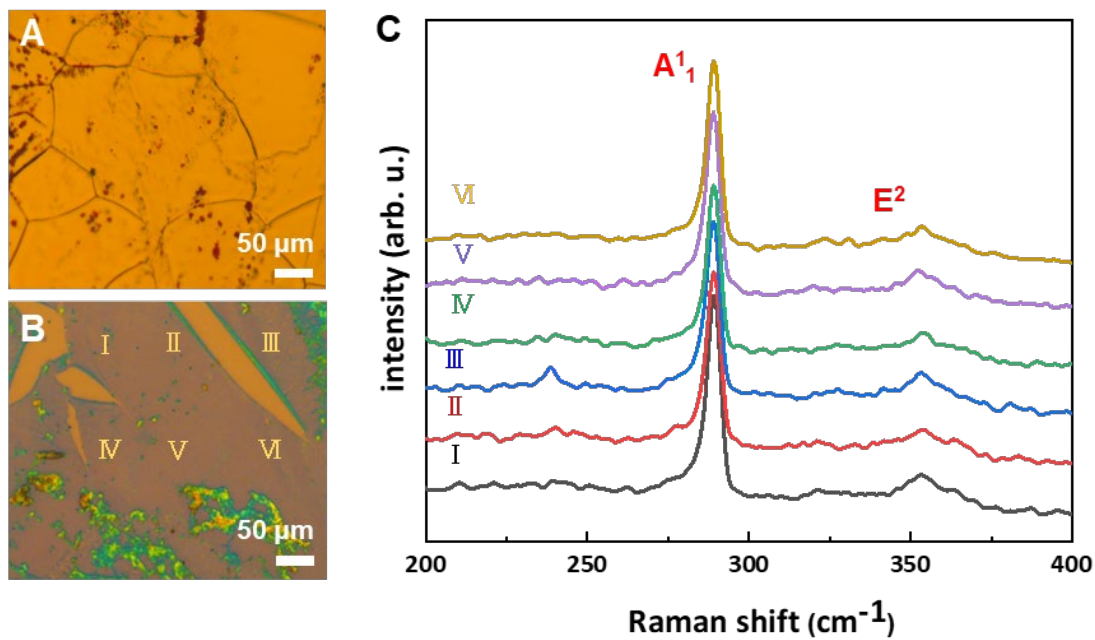

**Fig. S9:** **(A)** Optical microscopy image (OM) of large-area Janus SeMoS continuous monolayer film on Au foil. The synthesis of Janus film starts with the growth of $MoSe_2$ continuous film on the Au foil at ~720 °C for ~30min. **(B)** OM of transferred sub-millimeter size Janus SeMoS films. **(C)** Raman spectra extracted from six different positions over the sub-millimeter sized region in **(B)**. The statistical study shows the uniformity of the large-area Janus SeMoS monolayer film. The narrow FWHM (FWHM ≈ 4.6 ± 0.2 $cm^{-1}$) assessed from these uniform spectra show the high quality of the Janus SeMoS films without the presence of alloying or amorphous regions on the film.

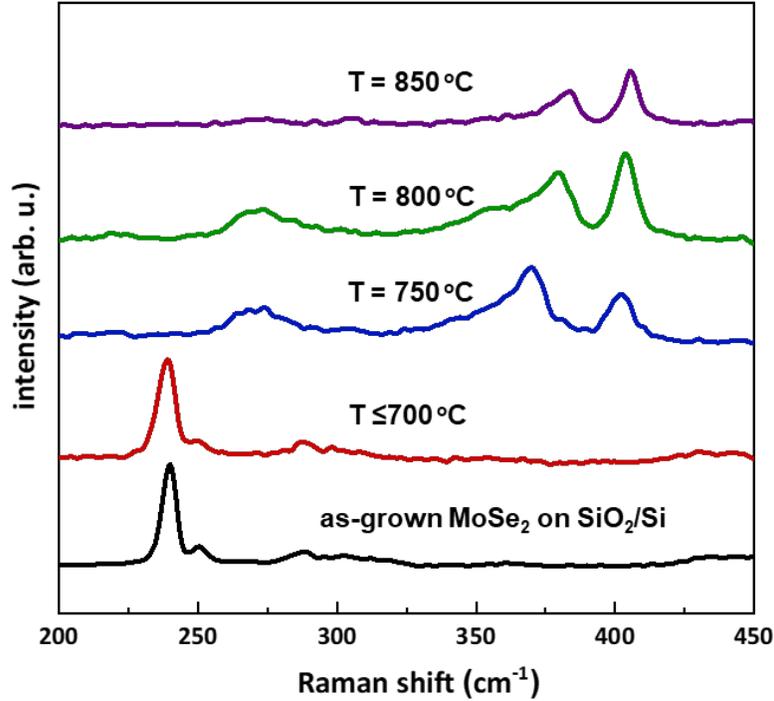

**Fig. S10: Influence of temperature in the sulfurization process of MoSe$_2$ ML on SiO$_2$/Si.** Raman spectra showing the influence of temperature in the sulfurization process of MoSe$_2$ ML on SiO$_2$/Si. All the growth and sulfurization processes performed together with the Au foils used for data presented in Fig. 1E for direct comparison between the growth on Au foils and SiO$_2$/Si substrates at different temperatures. Monolayer MoSe$_2$ crystals were first grown on both Au foils and SiO$_2$/Si substrates in each experiment followed by sulfurization for 10 minutes at different temperatures. Very importantly, no Janus SeMoS Raman signature observed on the crystals grown and sulfurized on SiO$_2$/Si substrate, while Janus SeMoS Raman signature is observed on the crystals grown and sulfurized on Au foils at 700 °C (for example Fig. S8). Furthermore, broader and weak Raman signature can be observed around 270 cm$^{-1}$ for sulfurization temperatures between 750 °C to 800 °C, corresponding to the formation of (MoS$_{2(1-x)}$Se$_{2x}$ alloy(*56*). The samples sulfurized at 850 °C shows characteristics peaks of MoS$_2$ with broader full width at half maximum (FWHM) indicating the conversion to highly defective MoS$_2$ (*48, 49*). When the sulfurization temperature below 700 °C, only MoSe$_2$ signature can be observed, which indicated that S does not substitute both of top and bottom Se layer of MoSe$_2$ monolayer without the catalytic effect of Au substrate. Thus, Au play an essential role during the Janus SeMoS structure formation and can act as catalyst to replace the bottom Se layer.

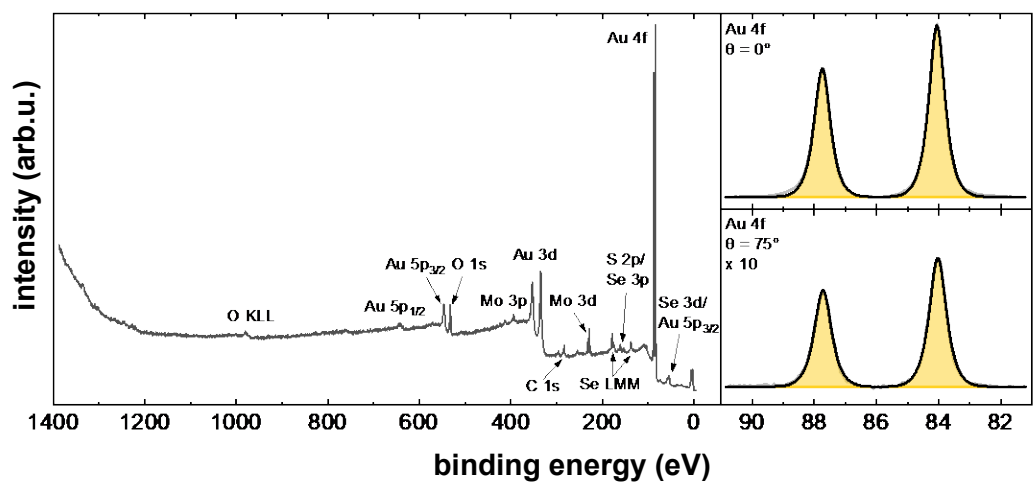

**Fig. S11.** (left) Survey XP spectrum taken at the emission angle with maximum intensity, where the main photoelectron orbital and Auger peaks are named. Besides Se, Mo, S and Au as substrate, C and O could also be measured as adsorbates from the ambient conditions. (right) Au 4f XP spectra at normal emission (θ = 0°, top) and at θ = 75° (right). The intensity of the spectrum at 75° were multiplied by the 10. These Au peaks are also used for the calculation in Fig. 2B.

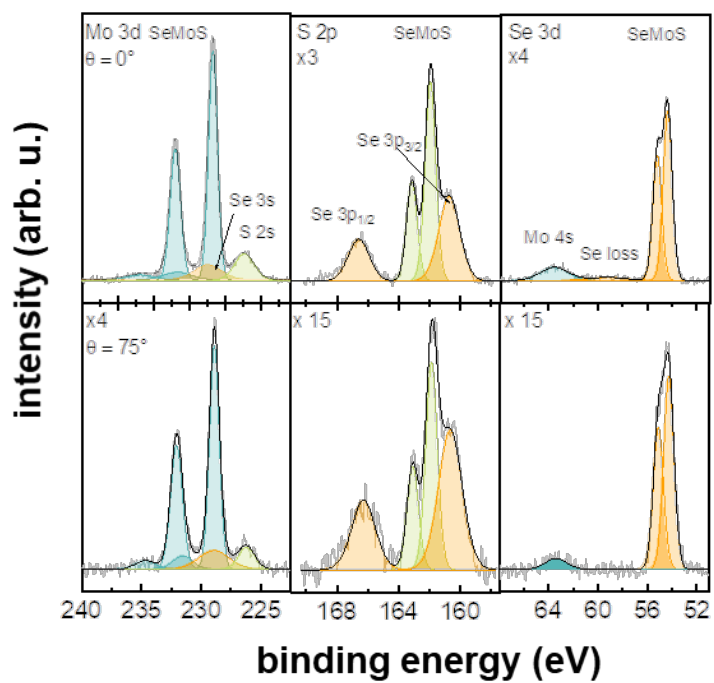

**Fig. S12.** High-resolution Mo 3d, S 2p, and Se 3d XP spectra of transferred Janus SeMoS on SiO$_2$/Si substrate measured at an emission angle (θ) of 0° (normal emission, top) and 75° (bottom), respectively. After transfer, we see the same SeMoS components in the spectra, which is an indication for a successful transfer. Also, we have the same conclusion, that Se layer should be on top, followed by Mo and S (same procedure as in Fig. 2 for calculation, but as substrate we used Si instead of Au). From the S 2p region we directly see this behavior for Se and S, because the Se 3p orbital increases in intensity compared to the S 2p orbital for higher emission angle. In contrast to the as grown Janus SeMoS sample, the Sulphur amount is reduced. The ratios Se:Mo:S are calculated to be 0.8±0.2:1.0±0.2:1.3±0.2.